\documentclass[prb,showpacs,double,twocolumn,10pt]{revtex4}
\usepackage{amsmath}
\usepackage{amssymb}
\usepackage{color}
\usepackage{graphics}
\usepackage{epsfig}

\begin{document}

\title{Kondo Insulator description of spin state transition in FeSb$_{2}$}
\author{$^{\ddagger }$C. Petrovic, $^{\ddagger \S }$Y. Lee, $^{\ddagger \S }$%
T. Vogt, $^{\P }$N. Dj. Lazarov, $^{\star }$S. L. Bud'ko and $^{\star }$P.
C. Canfield}
\affiliation{$^{\ddagger }$Physics Department, Brookhaven National Laboratory, Upton, New
York 11973-5000\\
$^{\S }$Center for Functional Nanomaterials, Brookhaven National Laboratory,
Upton, New York 11973-5000\\
$^{\P }$Laboratory of Theoretical and Solid State Physics, Institute of
Nuclear Sciences - Vinca, P. O. Box 522, 11001 Belgrade, Serbia \\
$^{\star }$Ames Laboratory and Department of Physics and Astronomy, Iowa
State University, Ames, Iowa 50011}
\date{\today }

\begin{abstract}
The thermal expansion and heat capacity of FeSb$_{2}$ at ambient pressure
agrees with a picture of a temperature induced spin state transition within
the Fe t$_{2g}$ multiplet. However, high pressure powder diffraction data
show no sign of a structural phase transition up to 7GPa. A bulk modulus
B=84(3)GPa has been extracted and the temperature dependence of the Gr\"{u}%
neisen parameter has been determined. We discuss here the relevance of a
Kondo insulator description for this material.
\end{abstract}

\maketitle

FeSb$_{2}$ is a small gap semiconductor whose magnetic properties strongly
resemble FeSi. The magnetic susceptibility of FeSi evolves from a high
temperature maximum to a flat van-Vleck like behavior as T$\rightarrow $0
with a Curie tail which is sample dependant and attributed to defects and
impurities. Aeppli and Fisk have pointed out that the underlying physics of
FeSi is similar to the physics of Kondo insulators with localized magnetic
moments.\cite{FiskAeppli} This picture was also supported by other
measurements as well.\cite{Mason}$^{,}$\cite{Mandrus} All Kondo insulator
materials are cubic with the exception of CeRhSb and CeNiSn and are all 
\textit{4f }intermetallics, except FeSi, RuAl$_{2}$ and Fe$_{2}$VAl.\cite%
{FiskAeppli}$^{,}$\cite{Basov}$^{,}$\cite{Nishino} A simple model that
explains their properties involves a flat f band hybridizing with a broad
conduction band with two electrons per unit cell. At T=0 the electrons
populate a lower hybridized band, and with the increase of T the electrons
start populating the higher band, resulting in a thermally activated Pauli
susceptibility. The Anderson Kondo lattice model with two electrons per site
with one \textit{f} and one conduction band thus captures the important
physics. Cubic structures are more favorable for this scenario since the
requirement of just one band crossing the Fermi level is more likely.

\begin{figure}
\epsfxsize=.45\textwidth
\epsfbox{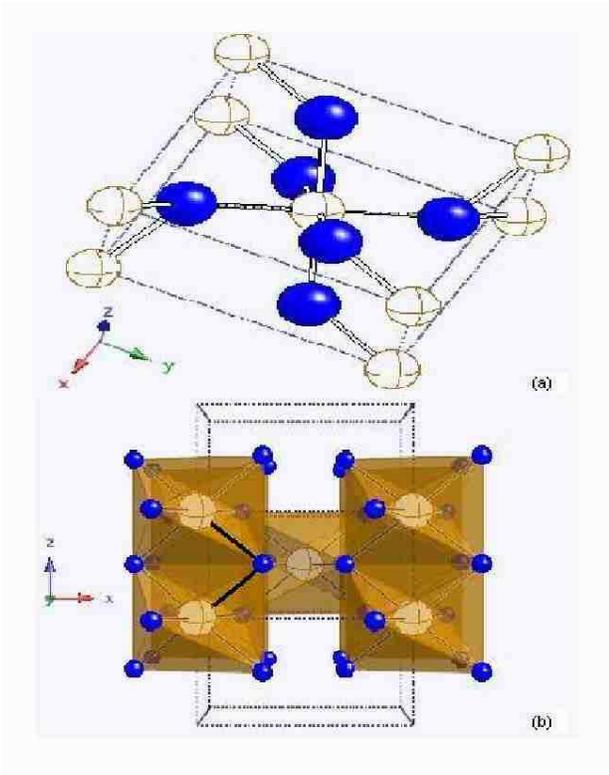}
\caption{(a) Crystal structure of FeSb$_{2}$ showing Fe (open, brown
crosses) surrounded by Sb (full blue) octahedra. The two short Fe-Sb bond
distances are shown in black and white. The d$_{xy}$ orbital is oriented in
between short Fe-Sb bond distances along the z-axis (the c-axis of the unit
cell). (b) The Fe-Sb-Fe bond angle $\protect\alpha $ associated with the
edge sharing octahedra along the c-axis of the unit cell. Note that
increased occupancy of d$_{xy}$ orbitals is reflected in the increase of $%
\protect\alpha $.}
\end{figure}

Another way to think about FeSi is based on the picture of a nearly magnetic
semiconductor in an itinerant model. \cite{Moriya}$^{,}$\cite{Edwards} This
picture was supported by local density approximation (LDA) band structure
calculations. \cite{Anisimov}$^{,}$\cite{Anisimov2} An important question to
answer is if the itinerant picture is correct or if the framework of the
Anderson model with a reduced on-site Coulomb repulsion U better describes
the physics of FeSi. Other possible model systems containing \textit{3d}
metals where these scenarios can be studied are therefore highly sought
after.

The small gap semiconductor FeSb$_{2}$ crystallizes in the marcasite-type FeS%
$_{2}$ structure (Fig. 1(a)).\cite{Nature198Hulliger1081} The basic
structural unit is made out of Fe ions surrounded by a deformed Sb octahedra.%
\cite{Hulliger} The Sb octahedra share edges along the \textit{c} axis (Fig
1(b)). According to "classical" ligand field theory, the Fe t$_{2g}$
orbitals are further split into two lower lying $\Lambda $ orbitals with d$%
_{yz}$ and d$_{xz}$ symmetries and higher lying $\Xi $ orbitals with d$_{xy}$
symmetry, which are directed toward near-neighbor cations along the \textit{c%
}-axis of the crystal.\cite{JSSC5Goodenough144} One type of marcasite the
so-called "anomalous marcasite" phase is formed with cations having d$^{n}$
configurations for n$\geq $6 and a large \textit{c}/\textit{a} ratio of
0.73-0.75 in which the d$_{xy}$ orbitals are doubly occupied. The other
so-called "regular marcasite" structure is formed \ with n$\leq $4 for
smaller \textit{c}/\textit{a} ratios of 0.53-0.57 because the d$_{xy}$
orbitals are empty.\cite{Hulliger}$^{-}$\cite{StructureFeSb2} Consequently,
the Fe-Sb-Fe bond angle $\alpha $ between the neighboring cations in
edge-sharing octahedra along the c-axis is $\alpha <90^{\circ}$ 
in regular and $\alpha > 90 ^{\circ }$ in anomalous
marcasites (see Fig. 1(b)).\cite{JSSC5Goodenough144} Thus, if the $d^{n}$
configuration of transition metal atom has $n > 4$, then it implies
electrons in $\Xi $ orbitals. Repulsive forces along the c-axis will
consequently increase the angle $\alpha $. For example, FeS$_{2}$ with a $%
d^{6}$ configuration has $\alpha $=97.5$^{\circ }$ as opposed to FeAs$_{2}$
with a $d^{4}$ configuration and $\alpha $=72.5$^{\circ }$. Therefore the
angle $\alpha $ should increase monotonically with the number of $\ \Xi $
electrons from a $d^{4}$ configuration ($\Xi $ empty) to a $d^{6}$
configuration ($\Xi $ filled).\cite{JSSC5Goodenough144}

\begin{figure}
\epsfxsize=.45\textwidth
\epsfbox{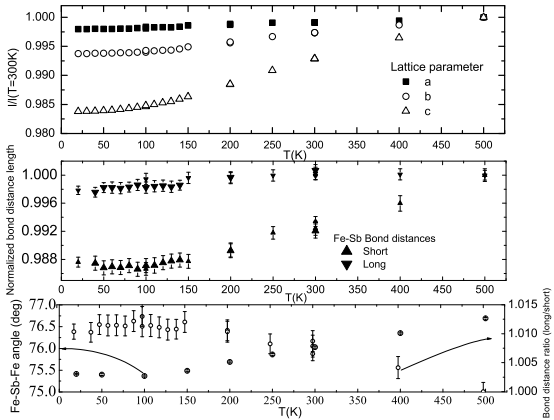}
\caption{Temperature dependance of (a) normalized unit cell parameters (b)\
Bond distance lengths and (c) Fe-Sb-Fe angle $\protect\alpha $ between edge
sharing octahedra and bond distance ratio (long/short). Note the increase of
angle $\protect\alpha $ above T$_{crossover}$ (see text) and the presence of
two regions of the Fe-Sb normalized short bond distance length contraction
around it.\ \ \ }
\end{figure}

The ionic configuration of iron in FeSb$_{2}$ is d$^{4}$ and its ground
state at low temperatures is the low spin state Fe$^{4+}$ (t$_{2g}^{4}$e$%
_{g}^{0}$) with the $\Lambda $ orbital filled and the $\Xi $ orbital empty.%
\cite{JSSC5Goodenough144} Indeed, at low temperature the magnetic
susceptibility of FeSb$_{2}$ results entirely from the core diamagnetism and
shows little temperature dependence.\cite{ja} Similarly to FeSi, above 100K
the susceptibility is paramagnetic.\cite{ja}$^{,}$\cite{JSSC5Fan136} One
possible explanation for the enhanced magnetic susceptibility is a spin
state transition of the Fe ions.\cite{ja} It is not uncommon that in
transition metal alloys the crystal field energy marginally exceeds the
Hund's exchange energy and therefore energies of the order of 1meV can
produce excitations across a spin gap. What makes anomalous marcasites
unusual is the existence of a band of itinerant electronic states due to the
overlap of higher lying $\Xi $ orbitals along the c-axis.\cite%
{JSSC5Goodenough144} In this work, structural, thermodynamic and magnetic
studies were carried out in an effort to further elucidate the
structure-property relation in this material since a spin state transition
within the $3d$ multiplet might couple to the lattice.

High quality FeSb$_{2}$ single crystals have been grown from excess Sb flux.%
\cite{ja} Variable temperature and high pressure experiments were performed
at the beamline X7A of the National Synchrotron Light Source at the
Brookhaven National Laboratory. A closed-cycle He-refrigerator and a
diamond-anvil cell were used to control the sample environment, and
monochromatic synchrotron X-ray and gas-proportional position-sensitive
detector were used to measure the powder diffraction data.\cite{X7A}
Magnetization and specific heat were measured in Quantum Design MPMS and
PPMS\ instruments. Single crystals of FeSb$_{2}$ were thoroughly ground to a
fine powder. Rietveld refinements were performed using GSAS\cite{GSAS}. The
results of fits were summarized in Tables 1 and 2.

\begin{figure}
\epsfxsize=.45\textwidth
\epsfbox{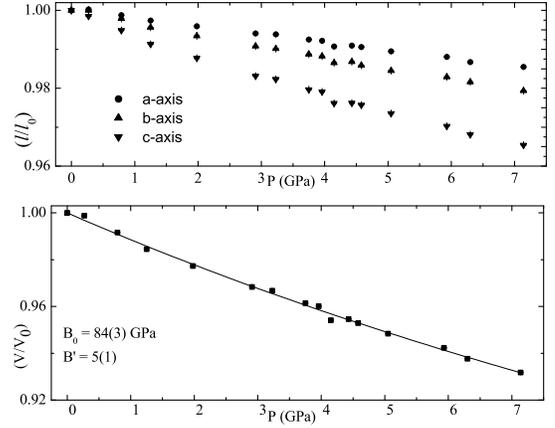}
\caption{Pressure dependance of (a) Normalized unit cell parameters (b) Unit
cell volume of FeSb$_{2}$}
\end{figure}

Fig. 2 shows the temperature evolution of several structural parameters.
Linear coefficients of thermal contraction\cite{Hazen} $\alpha $ in the
paramagnetic state\cite{ja} $\alpha _{i}=1/l(\partial l/\partial T)_{P}$ (i
= a, b, c) for individual unit cell axes are: $\alpha _{a}=4.2\cdot
10^{-6}K^{-1},\alpha _{b}=1.3\cdot 10^{-5}K^{-1}$and $\alpha _{c}=3.4\cdot
10^{-5}K^{-1}$. Individual Fe-Sb bonds also change as the temperature is
lowered and temperature effects on bond distance lengths normalized to 500K
are shown in Fig. 2(b). We observe two regions of bond contraction with a
crossover near 100K. The thermal contraction of the long Fe-Sb bond
distances (in the octahedral plane)\ becomes marginally stronger below 300K,
whereas the short (apical) Fe-Sb bond distances contract significantly
between 500K and 100K. The contraction of the short Fe-Sb bond distances is
finished at 100K and below this temperature we observe no further thermal
contraction on cooling down to the lowest measured temperature of 20K. We
can estimate the bond distance contraction by calculating the mean
coefficient of contraction between 500K\ and 20K:\ $\overline{\alpha }%
_{T_{1},T_{2}}=2/(i_{1}+i_{2})\left[ \frac{i_{1}-i_{2}}{T_{2}-T_{1}}\right] $
(T$_{1}$=20K, T$_{2}$=500K; i=short,long): $\overline{\alpha }%
_{20,500}^{short}=1.16\cdot 10^{-5}K^{-1},\overline{\alpha }%
_{20,500}^{long}=2.6\cdot 10^{-5}K^{-1}$, indicating relatively large
differences in the distortion of the octahedra. Both the bond distance ratio
and the Fe-Sb-Fe angle $\alpha $ associated with the shared octahedral-site
edges along the c-axis (Fig. 2 (c)) change the slope below 100K. The change
in the bond distance ratio implies maximum distortions for the Fe-Sb
octahedra around the T$_{crossover}\sim $100K. The value of the Fe-Sb-Fe
bond angle at 300K (76.028$\pm $0.02)$^{\circ }$ agrees well with the
previously reported value of 76.0$^{\circ }$.\cite{Brostigen}$^{,}$\cite%
{StructureFeSb2} Fig. 2(c) shows that the Fe-Sb-Fe bond angle $\alpha $ has
little or no temperature dependance for T < 100K and increases
above T$_{crossover}\sim $100K.

The lattice parameters contract smoothly with pressure to smaller values
above 3GPa than the ones obtained by thermal contraction down to 20K (Fig.
3(a), Fig. 4). Individual lattice parameter compressibilities $\beta
_{0}^{i}=-\frac{1}{i}\left( \frac{\partial i}{\partial P}\right) _{T}$ $%
(i=a,b,c)$ are quite different:\ $\beta _{0}^{a}=0.00223(4)GPa^{-1}$, \ $%
\beta _{0}^{b}=0.00339(3)GPa^{-1}$and \ $\beta _{0}^{c}=0.00655(5)GPa^{-1}$.
The ratio of the \textit{c} and \textit{a} axis compressibilities is as much
as $(\frac{\beta _{c}}{\beta _{a}})_{T}$=2.9. When compared with a ratio of
thermal contraction $(\frac{\alpha _{c}}{\alpha _{a}})_{P}$=8, it follows
that structural changes are more anisotropic with temperature. Both
temperature and pressure compress the \textit{c}-axis which is the direction
of higher lying $\Xi $ orbitals associated with d$_{xy}$ symmetry rather
strongly, whereas the \textit{b} and \textit{a} axis are less compressible.
The ratios of contraction to compressibility are also highest for the
c-axis: $\left( \frac{\alpha }{\beta }\right) ^{a}=1.88\cdot
10^{-3}GPaK^{-1} $, $\left( \frac{\alpha }{\beta }\right) ^{b}=3.83\cdot
10^{-3}GPaK^{-1}$ and $\left( \frac{\alpha }{\beta }\right) ^{c}=1.15\cdot
10^{-2}GPaK^{-1}$. We see no evidence for the formation of a high pressure
polymorph up to (7.14$\pm $0.1)GPa, as was established in the cases of CrSb$%
_{2}$ and NiSb$_{2}$.\cite{TakizawaPSS} The uncertainty in Rietveld
refinement and the errors of the bond distances at high pressures limit our
conclusions on the bond distance ratio and the Fe-Sb-Fe octahedral angle
between neighboring cations in edge-sharing octahedra. However, preliminary
measurements indicate that there is no change in the pressure dependence of
the Fe-Sb bond distances and the angle $\alpha $ in contrast to their
thermal contraction. Their ratio remains constant under pressure. Fig. 3 (b)
shows the normalized unit cell volume as a function of pressure. We obtain
the value of bulk modulus B$_{0}$=84(3)GPa and B$^{^{\prime }}$=5(1) by
fitting a Birch-Murnagham equation of state.\cite{BM}

The Fe-Sb long and Fe-Sb short bond lengths in Fe-Sb octahedra involve Fe a$%
_{\sigma }$ bonding orbitals (e$_{g}$ in origin) and Sb sp$^{3}$ hybrid
orbitals, as pointed out by Goodenough and Fan.\cite{JSSC5Goodenough144}$^{,}
$\cite{JSSC5Fan136} Bond distance lengths in Fe-Sb octahedra influence the
large energy gap between t$_{2g}$ and e$_{g}$ derived states. The gap within
the t$_{2g}$ multiplet is caused by mixing of $\Xi $ cation and $\sigma $
bonding anion orbitals, and the strength of this mixing is governed by the
Fe-Sb-Fe angle $\alpha $ that measures the deviation from an ideal
tetrahedral angle.\cite{JSSC5Goodenough144} Starting from the diamagnetic
state state at low temperatures, the increase of the the Fe-Sb-Fe angle $%
\alpha $ between neighboring cations (Fig. 1(b), Fig. 2(c)) above T$%
_{crossover}$ $\sim $100K indicates a thermally induced population of higher
lying $\Xi $ orbitals (d$_{xy}$ in origin). Since T$_{crossover}$ coincides
with the temperature of diamagnetic to paramagnetic transition in FeSb$_{2}$%
, we conclude that the spin state transition observed in Ref. 14 occurs
within the Fe t$_{2g}$ multiplet. \cite{ja}

\begin{figure}
\epsfxsize=.45\textwidth
\epsfbox{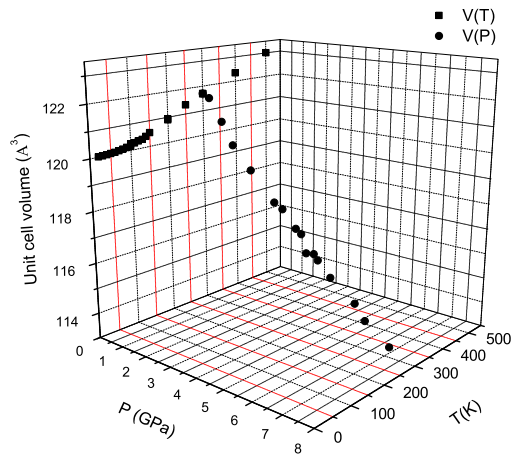}
\caption{The effects of pressure and temperature on the unit cell volume.
The 8GPa of pressure reduce the unit cell volume by 1.4\%, 4.6 times
exceeding the ambient pressure compression for 6.5\% by temperature decrease
from 300K to 20K.}
\end{figure}

Goodenough's description of the energy bands of FeSb$_{2}$ can be connected
with the electronic and magnetic properties of Kondo insulators. Transitions
within the t$_{2g}$ multiplet involves the population of a d$_{xy}$-derived
band of itinerant states. In the case of a Kondo insulator we expect to
describe the magnetic susceptibility with a model of a metallic spin
paramagnetism, albeit with a large renormalization of the noninteracting
bands. Thus, we used the model proposed by Jaccarino.\cite{Jaccarino} The
temperature dependence of the magnetic susceptibility should be explained
with a model of two narrow bands (peaks) at the density of states of width W
and separated by E$_{g}$=2$\Delta $ (see inset in Fig. 5). This model has
been successfully applied to show the validity of Kondo insulator picture in
seminal work on the thermodynamics of FeSi.\cite{Mandrus}

The Pauli susceptibility of an itinerant electron system with N(E) density
of states is:

\begin{equation}\nonumber
\chi (T)=-2\mu _{B}^{2}\underset{c-band}{\int }N(E)\frac{\partial f(E,\mu,T)}{\partial E}dE
\end{equation}

where $\mu _{B}$ is the Bohr magneton \ and $f(E,\mu ,T)=(\exp [(E-\mu
)/k_{B}T]+1)^{-1}$. The factor 2 in the above equation is due to the holes
in the valence band that contribute to $\chi $ in the same way as electrons
in the conduction band.\cite{Paschen} Taking $N(E)=Np/W$ and $\mu =W+\Delta $%
, where $p$ is the number of states/cell, and N number of unit cells, we
obtain:
\begin{equation}\nonumber
\chi (T)=-2\mu _{B}^{2}\frac{Np}{W}\frac{\exp (\beta \Delta )(1-\exp (\beta
W))}{(1+\exp (\beta \Delta ))(1+\exp (\beta (\Delta +W))}+\chi_0
\end{equation}
By fitting magnetic susceptibility data we obtain allowable values for
parameters: $0K<W/k_{B}<400K$ and $850K <Eg/k_{B}<1100K$, 
comparable to parameter values obtained for
FeSi.\cite{Mandrus}\qquad 

\begin{figure}
\epsfxsize=.45\textwidth
\epsfbox{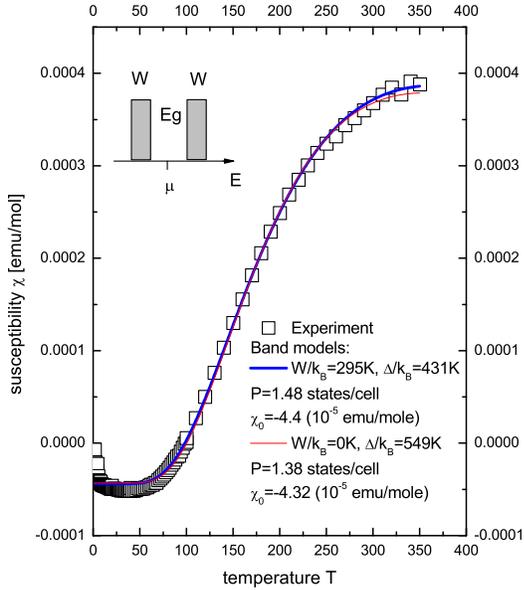}
\caption{Magnetic susceptibility of FeSb$_{2}$ (data taken from Ref. 14).
Note much smaller low temperature Curie tail than in FeSi. Solid and dashed
lines represent fits using the model density of states shown in the inset.}
\end{figure}

\begin{figure}
\epsfxsize=.45\textwidth
\epsfbox{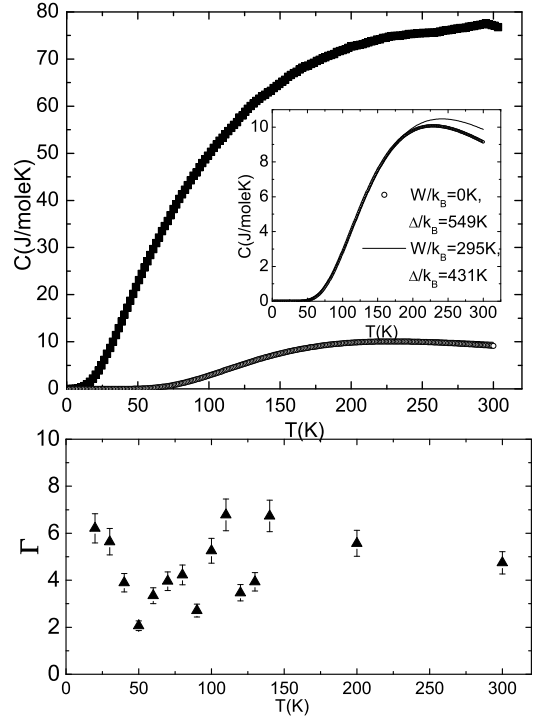}
\caption{(a) Heat capacity C$_{P}$ of FeSb$_{2}$. Inset shows contribution
to heat capacity from the model density of states used to model magnetic
susceptibility (see text). (b) Gr\"{u}neisen parameter $\Gamma $ of FeSb$_{2}
$}
\end{figure}

\begin{figure}
\epsfxsize=.45\textwidth
\epsfbox{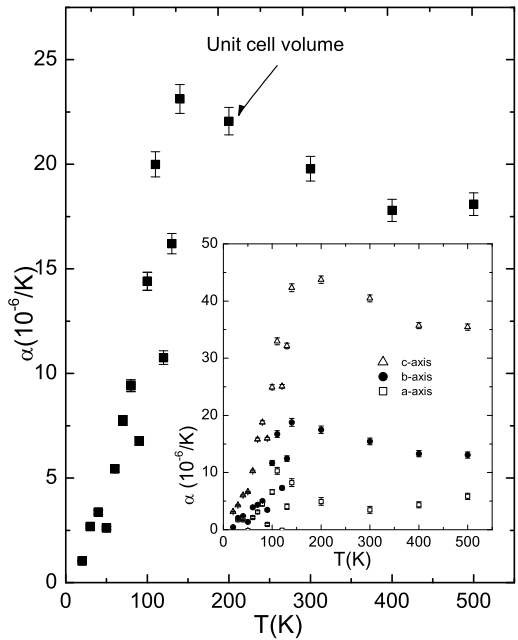}
\caption{Thermal expansion $\protect\alpha =\frac{1}{V}(\frac{\partial V}{%
\partial P})_{P\text{ }}$ of the unit cell volume of FeSb$_{2}$. Weak
temperature dependance of the unit cell volume introduces some noise in
thermal expansion data below 100K. Inset shows thermal expansion coefficient
of the individual axes $\protect\alpha =\frac{1}{l}(\frac{\partial l}{%
\partial P})_{P\text{ }}$(l=a,b,c)}
\end{figure}

The accumulation of states at the gap edges should in principle be observed
in other thermodynamic properties, such as the thermal expansion $\alpha $
and the heat capacity C$_{P}$. Unfortunately, there is no \textit{MSb}$_{2}$
diamagnetic member with M=\textit{3d} transition metal element with the
marcasite structure that we could use in order to extract the electronic
contribution to $\alpha $ and C$_{P}$. A similar difference has been
obtained for FeSi using CoSi as a reference.\cite{Mandrus} CoSb$_{2}$ forms
in a distorted monoclinic structure $P2_{1}/c$ at ambient temperature and
NiSb$_{2}$, contrary to previously reported, is not diamagnetic.\cite%
{Holseth} From the low temperature specific heat (Fig. 6a) $C\sim BT^{3}$
where $B=\frac{12}{5}nR\pi ^{4}\left( \frac{T}{\theta _{D}}\right) ^{3}$, we
estimate the Debye temperature as $\theta _{D}=(256\pm 6)K$, a smaller value
than previously reported ($\theta _{D}=380K$) based on vapor transport grown
samples. Using the same range of parameters obtained from the magnetization
fit, we have calculated spin state transition contribution to specific heat C%
$_{P}$=($\partial U/\partial T$) with U given by the contributions from
valence and conduction band:

\begin{eqnarray}\nonumber
U &=& \underset{0}{\overset{W}{\int }}\frac{Np}{W}\frac{EdE}{\exp (\beta
(E-W-\Delta ))+1} \cr\cr
&& \hskip .1in + \overset{2W+2\Delta }{\underset{W+2\Delta }{\int \frac{Np%
}{W}}}\frac{EdE}{\exp (\beta (E-W-\Delta ))+1}.
\end{eqnarray}
The results are shown in Fig. 6a where we see that phonon contribution
dominates at T$\sim $T$_{crossover}$. From the measured specific heat (Fig.
6a) and the unit cell volume temperature dependance (Fig. 4) we estimate the
Gr\"{u}neisen parameter $\Gamma =3B\alpha /C_{P}$ (Fig. 6b), where $\alpha =%
\frac{1}{3V}(\frac{\partial V}{\partial P})_{P\text{ }}$(Fig. 7) and B is
the bulk modulus. The overall magnitude of the thermodynamic Gr\"{u}neisen
parameter $\Gamma $ is comparable to the one seen in FeSi.\cite{Vocadlo} We
see that $\Gamma $ is weakly temperature dependant with a possible minimum
developing around T=50K.\ Since $\Gamma $ represents the measure of
anharmonic effects in the lattice potential energy, our $\Gamma (T)$
indicates the overall agreement with the Gr\"{u}neisen assumption $\Delta
\omega /\omega =-\Gamma \Delta V/V$.

The $\alpha (T)$ for FeSb$_{2}$ (Fig. 7) is different from the case of cubic
or anisotropic metals, where $\alpha (T)$ follows the C$_{P}$ decrease below
room temperature in accord with the Debye function. Departures are expected
for large variations of $\Gamma $.\cite{BarronWhite} The electronic
contribution to the specific heat of a narrow gap semiconductor is expected
to show a Schottky peak, as expected for a two level electronic system
separated by gap E$_{g}$=2$\Delta $. If the degeneracies of levels separated
by E$_{g}$ are equal, we expect the peak at a temperature equivalent to $%
\Delta =2.4T_{peak}$.\cite{MadrusTE}$^{,}$\cite{Aronzon} While we cannot
extract the electronic contribution from the phonons in C$_{P}$ (Fig. 6a
shows that it is $\sim $10\% of the measured value of C$_{P}$), since the Gr%
\"{u}neisen parameter $\Gamma =3B\alpha /C_{P}$ remains roughly constant,
the Schottky term should be reflected in the thermal conductivity as well.
Notwithstanding that we were not able to extract the electronic contribution
to $\alpha $, the data at Fig. 7 could be consistent with a broad peak
forming at T$_{peak}\sim $(125-200)K, thus giving $\Delta /k_{B}\sim
(300-480)K$. This might indicate that physically meaningful values of the gap
are $850K < E_{g}/k_{B} < 960K $, closer to the lower
limit of those obtained from the fit of the magnetic susceptibility.
Anisotropic thermal expansion coefficients of the the lattice constants $%
\alpha _{l}=\frac{1}{l}(\frac{\partial l}{\partial P})_{P\text{ }}(l=a,b,c)$
(Fig. 7 inset) show that the peak in the volume thermal expansion is
dominated by the c-axis contribution $\alpha _{c}$, whereas we see little or
no evidence for the peak formation for $\alpha _{a\text{ }}$or $\alpha _{b}$%
. Since the d$_{xy}$ orbitals that could be involved in a spin state
transition within the t$_{2g}$ multiplet overlap along the c-axis of the
crystal\cite{JSSC5Goodenough144}, it is likely that the electronic states
involved in the Schottky peak formation are in fact of d$_{xy}$ origin. This
agrees with Goodenough's picture of a band of itinerant states within the t$%
_{2g}$ multiplet and observations of enhanced conductivity in paramagnetic
region.\cite{JSSC5Goodenough144}$^{,}$\cite{ja}$^{,}$\cite{JSSC5Fan136}

In conclusion,we have shown that the experimental data for FeSb$_{2}$ can be
viewed as a thermally induced spin state transition within the Fe \textit{t}$%
_{2g}$ multiplet. A simple model with two peaks at the density of states at
the gap edges can explain the thermally induced paramagnetic moment, as
expected for Kondo Insulators. The magnetic properties correlate with the
electronic conductivity in FeSb$_{2}$ since the resistivity in the low spin
diamagnetic state at 2K is more than four orders of magnitude larger than in
the paramagnetic high spin state at room temperature.\cite{ja} This
significant enhancement of the conductivity in the paramagnetic state agrees
with Goodenough's hypothesis that thermal excitation populates $\Xi $
orbitals which have a substantial degree of covalent mixing and itineracy,
as opposed to the more localized $\Lambda $ orbitals.\cite%
{JSSC5Goodenough144}\ \ We see, however, that the magnetic susceptibility
can be described by both a thermally induced Pauli susceptibility (Fig. 5)
and a low spin to high spin transition.\cite{ja} Neutron scattering
experiments would be very useful to quantify this and to correlate the
phonon structure and frequency change with the thermally induced volume
change. We conclude that the FeSb$_{2}$ is a promising model system to study
the applicability of the Kondo Insulator vs. the nearly itinerant magnetic
semiconductor picture for \textit{3d} intermetallic compounds, in addition
to FeSi. This is further supported by preliminary optical spectroscopy
measurements.\cite{Leo}

We thank Zachary Fisk, T. M. Rice and Igor Zaliznyak for useful
communication. This work was carried out at the Brookhaven National
Laboratory, which is operated for the U.S. Department of Energy by
Brookhaven Science Associates (DE-Ac02-98CH10886) and at Ames Laboratory
which is operated for the U.S. Department of Energy by the Iowa State
University under Contract No. W-7405-82. This work was supported by the
Office of Basic Energy Sciences of the U.S. Department of Energy and in part
by the Serbian Ministry of Science and Technology through research grant
1899.

\begin{table*}[tbph]
\caption{FeSb$_{2}$ variable temperature refinements. Space group \textit{%
Pnnm}, Fe atom at \textit{2a} site at \textit{(000)}, Sb atom at \textit{4g}
site at \textit{(xy0)}, sample contains $\sim 8$wt.\% of Sb impurity. Other
values are available upon request}%
\begin{tabular}{|c|c|c|c|c|c|c|c|c|}
\hline\hline
T(K) & 20 & 100 & 150 & 200 & 250 & 300 & 400 & 500 \\ \hline\hline
a($\overset{\circ }{A}$) & 5.82117(6) & 5.82333(11) & 5.82480(8) & 5.82626(8)
& 5.82755(8) & 5.82764(9) & 5.82957(8) & 5.83298(7) \\ \hline\hline
b($\overset{\circ }{A}$) & 6.50987(7) & 6.51336(12) & 6.51738(9) & 6.52304(9
& 6.52911(9) & 6.53345(10) & 6.54221(9) & 6.55077(8) \\ \hline\hline
c($\overset{\circ }{A}$) & 3.16707(4) & 3.17035(6) & 3.17525(5) & 3.18218(5)
& 3.18977(5) & 3.19630(5) & 3.20785(5) & 3.21926(4) \\ \hline\hline
V($\overset{\circ }{A^{3}}$) & 120.016(2) & 120.249(5) & 120.540(3) & 
120.939(3) & 121.367(3) & 121.698(4) & 122.342(3) & 123.010(3) \\ 
\hline\hline
Sb(x) & 0.1875(1) & 0.1872(2) & 0.1875(2) & 0.1882(2) & 0.1882(2) & 0.1887(2)
& 0.1894(2) & 0.1901(2) \\ \hline\hline
Sb(y) & 0.3554(1) & 0.3549(2) & 0.3550(2) & 0.3551(2) & 0.3558(2) & 0.3557(2)
& 0.3564(2) & 0.3574(1) \\ \hline\hline
$\chi ^{2}$ & 6.740 & 9.606 & 8.822 & 7.671 & 8.842 & 9.145 & 8.186 & 7.671
\\ \hline\hline
Fe-Sb $\times $ 2 ($\overset{\circ }{A}$) & 2.5583(7) & 2.5558(9) & 
2.5586(10) & 2.5628(10) & 2.5691(9) & 2.5708(10) & 2.5800(11) & 2.5904(9) \\ 
\hline\hline
Fe-Sb $\times $ 4 ($\overset{\circ }{A}$) & 2.5889(6) & 2.5931(8) & 2.5935(8)
& 2.5934(8) & 2.5944(8) & 2.5951(9) & 2.5948(8) & 2.5945(7) \\ \hline\hline
Fe-Sb-Fe ($\circ $) & 128.95(2) & 128.99(3) & 128.99(3) & 128.98(3) & 
128.83(3) & 128.86(3) & 128.74(3) & 128.58(3) \\ \hline
\end{tabular}%
\end{table*}

\begin{table*}[tbph]
\caption{FeSb$_{2}$ high pressure refinements. Space group P$_{nnm}$, Fe
atom at \textit{2a }site at \textit{(000)}, Sb atom at 4g site at \textit{%
(xy0)}, sample contains $\sim 8$wt.\% of Sb impurity}%
\begin{tabular}{|c|c|c|c|c|c|c|c|c|}
\hline\hline
P (GPa) & ambient & 0.79 & 1.98 & 2.91 & 3.75 & 4.15 & 5.93 & 7.14 \\ 
\hline\hline
a (\AA ) & 5.82788(9) & 5.8207(2) & 5.8042(2) & 5.7933(2) & 5.7843(2) & 
5.7740(2) & 5.7583(2) & 5.7432(3) \\ \hline\hline
b (\AA ) & 6.53336(9) & 6.5198(2) & 6.4907(2) & 6.4733(2) & 6.4599(2) & 
6.4456(2) & 6.4217(2) & 6.3986(3) \\ \hline\hline
c (\AA ) & 3.19626(5) & 3.1799(1) & 3.1570(1) & 3.1423(1) & 3.1312(2) & 
3.1201(1) & 3.1012(1) & 3.0857(2) \\ \hline\hline
V (\AA 3) & 121.699(4) & 120.676(6) & 118.933(6) & 117.843(5) & 117.001(6) & 
116.118(4) & 114.677(5) & 113.394(6) \\ \hline\hline
Sb(x) & 0.1887(1) & 0.1858(4) & 0.1854(4) & 0.1857(3 & 0.1839(4) & 0.1850(3)
& 0.1842(4) & 0.1842(4) \\ \hline\hline
Sb(y) & 0.3561(1) & 0.3552(4) & 0.3537(3) & 0.3518(3) & 0.3527(3) & 0.3558(3)
& 0.3585(4) & 0.3502(4) \\ \hline\hline
$\chi ^{2}$ & 6.681 & 2.625 & 2.394 & 6.126 & 4.001 & 19.91 & 22.06 & 12.89
\\ \hline\hline
Fe-Sb $\times $ 2 (\AA ) & 2.5733(7) & 2.556(2 & 2.535(2) & 2.519(2) & 
2.515(2) & 2.530(2) & 2.535(2) & 2.478(2) \\ \hline\hline
Fe-Sb $\times $ 4 (\AA ) & 2.5942(5) & 2.601(2) & 2.594(2) & 2.589(2) & 
2.588(2) & 2.570(1) & 2.557(2) & 2.567(2) \\ \hline\hline
Fe-Sb-Fe ($%
{{}^\circ}%
$) & 128.79(2) & 128.79(7) & 129.07(6) & 129.45(6) & 129.19(7) & 128.80(6) & 
128.35(6) & 129.75(7) \\ \hline
\end{tabular}%
\end{table*}

\end{document}